\documentclass[conference]{IEEEtran}
\IEEEoverridecommandlockouts   
 
\usepackage{cite}
\usepackage{float}
\usepackage{multicol}
\usepackage{multirow}

\usepackage{hyperref}
\usepackage{graphicx}
\usepackage{amsfonts}
\usepackage{amsmath}
\usepackage{color}
\usepackage{epstopdf}
\usepackage{subcaption}
\usepackage{multirow}
\usepackage{bbm}
\usepackage{bm}
\usepackage{comment}

\usepackage{forest}

\usepackage[ruled]{algorithm2e}

\title{ 
Enhancing  Distribution Resilience with Mobile Energy Storage: A Progressive Hedging Approach
}

\author{\parbox{6 in}{\centering Jip Kim,~\IEEEmembership{Student Member,~IEEE,} and Yury Dvorkin,~\IEEEmembership{Member,~IEEE,}\\
		\tt\small Email: \{jipkim, dvorkin\}@nyu.edu}
		
		\thanks{ The authors are with the Department of Electrical and Computer Engineering, Tandon School of Engineering, New York University. This work was supported in part by the US NSF 
 Grant No. ECCS-1760540.}
}
\SetKwRepeat{Do}{do}{while}
\begin{document}
	\maketitle
	\thispagestyle{empty}
	\pagestyle{empty}
	\begin{abstract}
	    Electrochemical energy storage (ES) units (e.g. batteries) have been field-validated as an efficient back-up resource that enhance resilience of the distribution system in case of natural disasters. However, using these units for resilience is not sufficient to economically justify their installation and, therefore, these units are often installed in locations where they incur the greatest economic value during  normal operations. Motivated by the recent progress in transportable ES technologies, i.e. ES units can be moved using public transportation routes, this paper proposes to use this spatial flexibility to bridge the gap between the economically optimal locations during normal operations and disaster-specific locations where extra back-up capacity  is necessary. We propose a two-stage optimization model that optimizes investments in mobile ES units in the first stage and can re-route the installed mobile ES units in the second stage to avoid the expected load shedding caused by disaster forecasts. Since the proposed model cannot be solved efficiently with  off-the-shelf solvers, even for relatively small instances,  we apply a progressive hedging algorithm. The proposed model and progressive hedging algorithm are tested through two illustrative examples on a 15-bus radial distribution test system.

	\end{abstract}


    \vspace{-0.35cm}
	\section{Introduction} \label{sec:intro}
	 \vspace{-0.15cm}
		The US power grid is vulnerable to natural disasters (e.g. flooding, extreme winds, earthquakes) as they increase the likelihood of critical equipment failures \cite{watson2014conceptual}.  Distribution systems are particularly affected by natural disasters due to the compounding effect of line outages, radial topology, and limited back-up resources.  Furthermore, large-scale disasters can affect the power grid in multiple locations causing the domino effect, which may in turn spread power outages across large geographical areas, even if some areas would not otherwise be affected by disasters \cite{neumayer2013assessing}. Therefore, it is important to contain the failures within distribution systems and prevent their further propagation in the transmission system.  
		
		Provided a timely disaster forecast, power grid operators can prevent these failures, or at least mitigate their impacts, by strategically placing flexible back-up resources in the distribution system to strengthen its weak nodes (buses) and weak links (lines) \cite{wang2016research}. Since the disaster forecasts are normally available on a short notice (for instance, NOAA forecasts are available 48-168 hours ahead \cite{rappaport2009advances}), only a few technologies are physically suitable to be deployed, relocated or activated within this time frame: distributed energy resources, including portable diesel generators \cite{lei2016mobile} and parked electric vehicles \cite{wang2016research}, adaptive microgrids \cite{wang2015self}, or preventive load shedding \cite{smith2014analysis}. The objective of this paper is to illustrate that mobile ES units, i.e. those that can be moved among different distribution system locations using regular transportation routes, are also economically and physically suitable for this application.
		
		A mobile ES unit, often referred to as `storage-on--wheels',  is an emerging technology that has been recently developed in the form of a trailer-mounted electrochemical battery. Consolidated Edison of New York is currently considering installing such mobile ES units to reduce the impact of PV generation on their  distribution system in  New York City  and defer costly distribution upgrades \cite{conEd}. As reported in \cite{conEd}, each mobile ES unit will feature a Lithium-ion battery that can store up to 800 kWh of energy with the maximum charging/discharging duration of 10 hours. Relative to other resilience resources, the mobile ES units have multiple advantages.  First, they are more environmentally friendly than portable diesel generators and can be used without unnecessary noise and air pollution. Furthermore, strict pollution standards in certain jurisdictions, especially metropolitan areas, prohibit the use of portable diesel generators during normal operations, which reduce their value to the distribution system. Second, unlike electric vehicles or adaptive microgrids, the mobile ES units can be directly operated by the power grid operator and do not require advanced communication infrastructure or engagement with electricity consumers. Finally, recent advances in material sciences are expected to enable mobile ES units with just-in-time re-configurable power and energy ratings (so-called flow batteries) that can further enhance the back-up flexibility that these units can provide in case of natural disasters. 
		 
		Since mobile ES units can be used during normal operations and for dealing with the effects of natural disasters, there is a need to integrate their mobility in planning tools for power grid operators. A typical state-of-the-art planning tool for stationary ES units, \cite{pandvzic2015near, 7587806}, is routinely formulated as a two-stage stochastic mixed-integer linear program (MILP) and considers ES units as stationary resources. In these tools, the first stage optimizes the ES locations and sizes, while the second stage fixes the first-stage decisions and co-optimizes the operation of existing resources and newly installed ES units. To immunize the first-stage investment decisions against a variety of operating conditions, the second stage normally considers multiple operating scenarios. If the ES units were mobile, the complexity of the planning tool would increase. First, the ES mobility implies that the ES location is not fixed in the second stage and needs to be optimized (so-called recourse decisions). Second, the recourse decision on each mobile ES unit is binary, where it attains the value of 1, if that unit needs to be moved, or 0 otherwise. The two-stage stochastic MILPs with binary recourse decisions are particularly computationally demanding, and often existing solution strategies, e.g. Benders' decomposition, perform poorly when applied to such problems \cite{Sherali2009}. This computational complexity can be overcome by using the progressive hedging (PH)  algorithm \cite{Watson:2010dn}. Recently, the PH algorithm has gained attention in the context of a two-stage stochastic unit commitment problem \cite{ryan2013toward}. Unlike Benders' decomposition, the PH algorithm does not exploit the two-stage structure of the underlying optimization and does not separate the first- and second-stage decisions. Instead, the PH algorithm partitions the original problem in a scenario-based fashion and the first- and second-stage decisions are optimized for each scenario independently. In each scenario-based problem, the relaxation of the first-stage decision is penalized with an exogenous penalty  coefficient. The algorithm iterates until the first-stage decisions across all scenarios converge with a given tolerance. The scenario-based decomposition used in the PH algorithm is shown to be an effective solution strategy for two-stage MILPs with binary recourse decisions and, therefore, it is applicable for planning tools with mobile ES units.
		
		The primary focus of this problem is two-fold:
		
		\begin{enumerate}
		    \item It takes the perspective of the distribution system operator and formulates the optimization problem to decide on the investments in mobile ES units. The proposed optimization is a two-stage mixed-integer program with binary recourse decisions, which account for the relocation of mobile ES units under each specific scenario. This optimization achieves the trade-off between the economic value of mobile ES units during normal operations and their ability to enhance the distribution system resilience in case of natural disasters.
		    \item The proposed optimization is solved using the PH algorithm. The numerical results demonstrate that this method outperforms off-the-shelf solvers in terms of the computational performance and solution accuracy. The numerical experiments also suggest that the PH performance  can be improved by tuning externalities (e.g. penalty coefficients). 
	
		\end{enumerate}
		
The remainder of this paper is organized as follows. Section~\ref{sec:formulation} and Section~\ref{sec:Solution_technique} describe the proposed optimization and the PH implementation. Section~\ref{sec:case} describes the case study and demonstrates the usefulness of mobile ES units. Section~\ref{sec:conclusion} concludes the paper.

	\section{Planning with Mobile ES units} \label{sec:formulation}
		In the following, we consider a distribution system layout typical to the US power sector. The system has a radial topology, where sets $\mathcal{B}$ and $\mathcal{L}$, indexed by $b$ and $l$, represent the distribution buses (nodes) and lines (edges). The single distribution system operator is responsible for all planning and operating decisions, as well as incurred costs. The distribution power flows are modeled using a second-order conic (SOC) relaxation of ac power flows that is computationally tractable, \cite{farivar2013branch}. In this case the objective of the system operator is to allocate the mobile ES units with fixed power and energy ratings (the set of units is denoted by $\mathcal{K}$, indexed by $k$) in such a way that these units are operated as stationary resources during the normal operations and can be moved to a different location in case of natural disasters. Set $\mathcal{S}$, indexed by $s$, contains scenarios of the normal operations and emergency conditions. Each scenario is assigned probability $\omega_s$ and has the number of time intervals denoted by set $\mathcal{T}$, indexed by $t$.
		
		The objective function of this optimization is then given as:
		\begin{align}
		&\min \big[ \gamma \!\cdot\!IC (x_k) +  \sum\limits_{s\in{\cal S}} {\omega}_s \!\cdot\! OC_s (u_{kbts}|x_k)\big] \label{objective_function}\\
		&IC (x_k)=\!\sum_{k \in {\cal K}} \left(C^\mathrm{P} \! \cdot \! \overline{P}_k \!+\! C^\mathrm{E} \!\cdot\! \overline{E}_k \right)\!\cdot\! x_k \label{IC}\\
			\begin{split}
				&OC_s (u_{kbts}|x_k)= \!\!\!\!\!  \sum_{t\in {\cal T} , b\in {\cal B}} \!\! C_b^{\mathrm{g}} \cdot p_{bts}^{\mathrm{g}} + \!\!\!\!\!\!\!\!\!\!\! \sum_{k \in {\cal K}, b\in {\cal B}, t \in {\cal T}} \!\!\!\!\!\!\!\!\! \!C_{b}^{\mathrm{VoLL}} \cdot p_{bts}^{\mathrm{ls}} \\& \ +  \!\!\!\!\!\!\!\!\!\sum_{k \in {\cal K}, b\in {\cal B}, t \in {\cal T}} \!\! \left(\left\vert \frac{h}{100}  \right\vert   C^{\mathrm{P}} \left( p_{kbts}^{\mathrm{ch}} + p_{kbts}^{\mathrm{dis}} \right) \right), \ \ \forall s \in {\cal S}\label{OC}
			\end{split}
		\end{align}
		Eq.~\eqref{objective_function} minimizes the sum of the investment cost $IC(\cdot)$, where $\gamma$ is a daily capital recovery factor from \cite{pandvzic2015near} that prorates the investment cost on a daily basis,  and the expected expected daily operating cost over all scenarios $OC_s (\cdot)$. Eq.~\eqref{IC} computes the investment cost of installing mobile ES unit $k$ with fixed power and energy ratings $\overline{P}_k$ and $\overline{E}_k$, each priced with parameters $C^\mathrm{P}$ and $C^\mathrm{E}$ as in \cite{pandvzic2015near, 7587806}. The installation decision is modelled by binary variable $x_k \in \big\{ 0,1 \big\}$. If $x_k =1$, mobile ES unit $k$ is installed, otherwise $x_k=0$. Note that  $\overline{P}_k$ and $\overline{E}_k$ are parameters, but can be extended to decision variables. Eq.~\eqref{OC} computes the operating cost for each scenario, which includes: operating cost of generation resources with the incremental cost $C_b^{\mathrm{g}}$ producing $p_{bts}^{\mathrm{g}}$, load shedding cost based on $p_{bts}^{\mathrm{ls}}$ and value of lost load $C_{b}^{\mathrm{VoLL}}$, and the ES degradation cost. The degradation cost is computed as explained in \cite{ortega2014optimal} and depends on charging ($p_{kbts}^{\mathrm{ch}}$) and discharging ($p_{kbts}^{\mathrm{dis}}$) decision variables and parameter $h$, the degradation slope taken from \cite{ortega2014optimal}.  As described below, $OC_s (u_{kbts}|x_k)$ depends on the binary decision $u_{kbts}\in\big\{0,1 \big\}$, which is a recourse decision on the placement of  mobile ES unit $k$ at bus $b$ at time interval $t$ under scenario $s$ and that in turn depends on the scenario- and time-independent decision $x_k$. The binary recourse is denoted in \eqref{objective_function} as $u_{kbts}|x_k$ and is implicitly internalized in the right-hand side of \eqref{OC} via variables $p_{kbts}^{\mathrm{ch}}$ and $p_{kbts}^{\mathrm{dis}}$.
        
        The mobile ES units are operated as ($\forall k \! \in \! {\cal K}, t \! \in \! {\cal T}, s \!\in\! {\cal S}$):
		\begin{align}
			\begin{split}
				&e_{kts}=e_{k,t-1,s} \!+ \! \sum_{b \in \cal{B}} \big(  p_{kbts}^{\mathrm{ch}} \!\! \cdot \! \aleph^{\mathrm{ch}}- p_{kbts}^{\mathrm{dis}}  / \aleph^{\mathrm{dis}} \big) , 
				\label{ESS_dynamic}
			\end{split}\\
			& 0 \le e_{kts} \le \overline{E}_k, 
			\label{ESS_capacity_limit}\\	
		& 0 \!\le\! p_{kbts}^{\mathrm{ch}} \!\cdot\! \aleph^{\mathrm{ch}} \!\le\! \overline{P} _k  \!\cdot\!  u_{kbts} ,\ \! \forall b \!\in\! {\cal B}\! 
		\label{charging_power_limit}\\
			\begin{split}
				& 0 \!\le\! p_{kbts}^{\mathrm{dis}} / \aleph^{\mathrm{dis}} \!\le\! \overline{P} _k  \!\cdot\! u_{kbts} ,\ \! \forall  b \!\in\! {\cal B}\! 
				\label{discharging_power_limit}\
			\end{split}\\
		& \!-\! K \!\cdot \!p_{kbts}^{\mathrm{ch}} \! \le \!q_{kbts}^{\mathrm{ch}} \!\le\! K \!\cdot\!  p_{kbts}^{\mathrm{ch}}  , \forall  b\! \in\! {\cal B}\!
		\label{ESS_PF_ch}\\
		& \!-\! K \!\cdot \!p_{kbts}^{\mathrm{dis}} \!\le \!q_{kbts}^{\mathrm{dis}} \!\le \! K \!\cdot \! p_{kbts}^{\mathrm{dis}}  , \forall b\! \in\! {\cal B}\! 
		\label{ESS_PF_disch}
		\end{align}
		Eq.~\eqref{ESS_dynamic} relates the energy state-of-charge  $e_{kts}$ and charging/discharging decisions with imperfect efficiency $\aleph^{\mathrm{ch}}   = \aleph^{\mathrm{dis}} <1$.  Eq.~\eqref{ESS_capacity_limit} limits the energy stored to $\overline{E}_k$. Eq.~\eqref{charging_power_limit}-\eqref{discharging_power_limit} limit the maximum charging and discharging power to $\overline{P} _k$ and binary variable $u_{kbts}\in\{0,1\}$ indicates if mobile ES unit $k$ is located at bus $b$ in time period $t$ and scenario $s$. If $u_{kbts} =0$, the energy state-of-charge remains unchanged. Eq.~\eqref{ESS_PF_ch} and \eqref{ESS_PF_disch} relate the reactive power injection of mobile ES units to their charging and discharging power via parameter $K$ given by a desirable power factor. Note that \eqref{charging_power_limit}-\eqref{discharging_power_limit} contain products of continuous and binary variables, which can be linearized using the big-M method (see \cite{7587806} for details).
	    
	    The distribution system is operated as ( $\forall  t \! \in \! {\cal T}, s \!\in\! {\cal S}$):
		\begin{align}		
			\begin{split}
				& \left( f_{lts}^{\mathrm{p}} \right)^2 \! +\! \left( f_{lts}^{\mathrm{q}}  \right)^2 \! \le \! \left((1\!\!-\!\alpha_{lts})S_l \right) ^2 \!, \ \  \forall l \! \in \! {\cal L}\label{LineLimit_FW}		
			\end{split}\\ 
			\begin{split}
				& \left( f_{lts}^{\mathrm{p}} \!\! - \!\! a_{lts} \!\cdot \! R_{l} \right)^2 \!\!\!+\!\! \left( f_{lts}^{\mathrm{q}} \!\! - \! a_{lts} \! \cdot \! X_l \right)^2 \!\! \le \!\! \left(( \! 1 \!\! - \! \alpha_{lts})S_l \right) ^2 \! ,  \forall l \! \in \!\! {\cal L}\label{LineLimit_BW}
			\end{split}\\
			\begin{split}
				& v_{s(l),t,s}-2\!\left(R_l \!\cdot\! f_{lts}^{\mathrm{p}} + X_l \!\cdot\! f_{lts}^{\mathrm{q}} \right)  + a_{lts} \!\left(R_l ^2 + X_l ^2 \right) \\&= v_{r(l),t,s},   \ \forall l \! \in \! {\cal L},\label{DistFlow1}
			\end{split}\\
				& \frac{\left(f_{lts}^{\mathrm{p}}\right)^2 + \left(f_{lts}^{\mathrm{q}} \right) ^2 }{a_{lts}} \le v_{s(l),t,s}, \ \forall l \! \in \! {\cal L}\label{DistFlow2}\\
			\begin{split}
				& \!\!- \!\!\!\!\!\!\!\sum_{l \vert r(l)=0 }\!\!\!\!\! \left( f_{lts}^{\mathrm{p}} \!\!\!-\! a_{lts} \!\!\cdot\! R_l \!\right) \!-\! p_{0,t,s}^{\mathrm{g}} \!\!+\! G_0 \!\cdot\! v_{0,t,s} \!=\! 0, \label{P_balance_root} 
			\end{split}\\
			\begin{split}
				& \!\!- \!\!\!\!\!\!\!\sum_{l \vert r(l)=0 }\!\!\!\!\! \left( f_{lts}^{\mathrm{q}} \!\!\!-\! a_{lts} \!\!\cdot\! X_l \!\right) \!-\! q_{0,t,s}^{\mathrm{g}} \!\!-\! B_0 \!\cdot\! v_{0,t,s} \!=\! 0,\label{Q_balance_root}
			\end{split}\\
			\begin{split}
				& f_{bts}^{\mathrm{p}} \!\!-\!\!\!\!\!\! \sum_{l \vert r(l) = b } \!\!\!\left( f_{lts}^{\mathrm{p}} \!\!-\! a_{lts} \!\cdot\! R_l \right) \!-\! p_{bts}^{\mathrm{g}} \!+\! P_{bts}^{\mathrm{d}} \!\!-\! p_{bts}^{\mathrm ls}\! \!+\! G_b \!\cdot\! v_{bts}
				\\&\!-\!\!\! \sum_{k \in {\cal K}} \!p_{kbts}^{\mathrm{dis}} +\!\! \sum_{k \in {\cal K}} \!p_{kbts}^{\mathrm{ch}} =\! 0, \forall b \!\in\! {\cal B}\!,\label{P_balance} 
			\end{split}\\
			\begin{split}
				& f_{bts}^{\mathrm{q}} \!\!-\!\!\!\!\!\! \sum_{l \vert r(l) = b } \!\!\!\left( f_{lts}^{\mathrm{q}} \!\!-\! a_{lts} \!\cdot\! X_l \right) \!-\! q_{bts}^{\mathrm{g}} \!+\! Q_{bts}^{\mathrm{d}} \!\!-\! q_{bts}^{\mathrm ls}\! \!-\! B_b \!\cdot\! v_{bts}
				\\&\!-\!\!\! \sum_{k \in {\cal K}} \!q_{kbts}^{\mathrm{dis}}+\!\! \sum_{k \in {\cal K}} \!q_{kbts}^{\mathrm{ch}}  =\! 0, \forall b \!\in\! {\cal B}\!,\label{Q_balance} 
			\end{split}\\
			&  \underline{P}_b^{\mathrm{g}} \le p_{bts}^{\mathrm{g}} \le \overline{P}_b^{\mathrm{g}}, \ \ \ \forall b \in {\cal B} ^{\mathrm{G}} , \label{PGbound}\\
			& \underline{Q}_b^{\mathrm{g}} \le q_{bts}^{\mathrm{g}} \le \overline{Q}_b^{\mathrm{g}}, \ \ \ \forall b \in {\cal B} ^{\mathrm{G}} ,\label{QGbound}\\
			& 0 \le p_{bts}^{\mathrm{ls}} \le {P}_{bts}^{\mathrm{d}}, \ \forall b \! \in \! {\cal B}, \label{PLSbound}\\
			& 0 \le q_{bts}^{\mathrm{ls}} \le {Q}_{bts}^{\mathrm{d}}, \ \forall b \! \in \! {\cal B},\label{QLSbound}\\
			& \underline{V}_b \le v_{bts} \le \overline{V}_b, \ \forall b \! \in \! {\cal B}, \label{Vbound}
		\end{align}
		Eq.~\eqref{LineLimit_FW}-\eqref{DistFlow2} are the SOC-based power flow model as in \cite{farivar2013branch}, where active and reactive power flows are $f_{lts}^{\mathrm{p}}$ and $f_{lts}^{\mathrm{q}}$ and squared magnitude of nodal voltages at sending and receiving buses of each line $l$ are $v_{s(l)}$ and $v_{r(l)}$. The apparent flow limit, resistance and reactance of each line are given by $S_{l}$, $R_{l}$ and $X_{l}$. Scenario-dependent parameter $0\leq \alpha_{lts} \leq 1$ is used to emulate the impact of natural disasters on the distribution system lines. If $ \alpha_{lts} =1 $, line $l$ is tripped. On the other hand, if $ 0 < \alpha_{lts} <1  $, line  $l$  is operated with  a reduced apparent flow limit. Finally, if $ \alpha_{lts} =0 $, line  $l$  is operated with the normal apparent flow limit. Eq.~\eqref{P_balance_root}-\eqref{Q_balance_root} enforce the active and reactive nodal power balance for the root bus (denoted with index $0$) of the distribution system. The nodal power balance for other distribution buses is enforced in \eqref{P_balance} and \eqref{Q_balance}, where $P_{bts}^{\text{d}}$ and $Q_{bts}^{\text{d}}$ are the active and reactive power demand. Active and reactive power injections of conventional generators are constrained in \eqref{PGbound} and \eqref{QGbound} using their minimum and maximum limits ($\underline{P}_b^{\mathrm{g}}$, $\underline{Q}_b^{\mathrm{g}}$, $\overline{P}_b^{\mathrm{g}}$, $\overline{Q}_b^{\mathrm{g}}$). The nodal load shedding is limited to the nodal demand in \eqref{PLSbound} and \eqref{QLSbound}. The nodal voltages are to be kept with the upper ($\overline{V}_b$) and lower ($\underline{V}_b$) limits enforced in \eqref{Vbound}. 
		
		The mobility of ES units is modeled as: 	
		\begin{align}	
			& \sum_{b \in {\cal B}} u_{kbts} \le x_k, \ \ \  \forall k \in {\cal K},  t \in {\cal T}, s \in {\cal S} \label{ESS_location} \\
			& \sum_{k \in {\cal K}} u_{kbts} \le N_b ^{\mathrm{ES}} , \ \ \  \forall b \in {\cal B} , t \in {\cal T}, s \in {\cal S}\label{CS_number}\\
			& u_{kbt,s_1} \! = \!u_{kb,t_1,s_1}, \  \forall s \!\in\! \mathcal{S}, k \! \in \! {\cal{K}}, b \! \in \! {\cal{B}}, t \! \in \! {\cal{T}} \backslash{\big\{t_1\big\}}\label{ESS_stationary} \\
			& u_{kb,t_0,s} \! = \!u_{kb,t_0,s_1}, \ \ \forall k \! \in \! {\cal K}, b \! \in \! {\cal B}, s \! \in \! {\cal{S}}\label{ESS_initial_location_link} \\
			\begin{split}
				& u_{kb_1ts}-u_{kb_1,t+1,s} \le 1 - u_{kb_2,t+\tau  ,s},\  \forall  s \in \mathcal{S} \backslash{\big\{s_1\big\}}, \\ 
				& k \!\in\! {\cal K}, b_1 \!\neq\! b_2 \!\in\! {\cal B}\!, t \!\in\! {\cal T}\!, \tau \!\in\! \left[1,\!\cdots\!,\min(T_{b1,b2,t}^{\mathrm{d}}, N^{\mathrm{t}} \!\! - \! t ) \right]\label{ESS_Delay}
			\end{split} 
		\end{align}
		Eq.~\eqref{ESS_location} relates $u_{kbts}$ to the investment decision  $x_k$ accounted for in \eqref{objective_function}. If $x_k=0$, it follows from \eqref{ESS_location} $u_{kbts}=0$. If $x_k=1$, $u_{kbts}$ is optimized for each scenario.  Eq.~\eqref{CS_number} limits the number of mobile ES units ($N_b ^{\mathrm{ES}}$) that can be connected to bus $b$. Eq.~\eqref{ESS_stationary} ensures that  mobile ES units are operated as stationary resources during the normal operations (denoted as scenario $s=1$).  Eq.~\eqref{ESS_Delay} models the transition delay on moving mobile ES unit from bus $b_1$ to bus $b_2$, where $T_{b_1, b_2, t}^{\mathrm{d}}$ is a given transition time between buses $b_1$ and $b_2$. In practice, the value of $T_{b_1, b_2, t}^{\mathrm{d}}$  can be determined based on the availability and length of transportation routes.   Note that \eqref{ESS_Delay} is somewhat equivalent of minimum up and down time constraints on the on/off status of conventional generators  modeled in UC problems, see \cite{ryan2013toward}.

		
	\vspace{-0.1cm}
	\section{Solution Technique}\label{sec:Solution_technique}
	The optimization problem in \eqref{objective_function}-\eqref{ESS_Delay} is a two-stage stochastic mixed-integer program with binary recourse decisions. In general, such problems can be solved using off-the-shelf solvers, but their performance is limited, especially for large networks.  To solve this problem efficiently, we apply the PH algorithm. Following \cite{Watson:2010dn}, we decompose the original problem in \eqref{objective_function}-\eqref{ESS_Delay} in $N^{\mathrm{s}}\!=\!card(\mathcal{S})$ subproblems, where each subproblem can be solved independently in a parallel fashion. Due to the pagination limit, we condense the traditional 9 steps of the PH algorithm, \cite{Watson:2010dn},  in three steps:
	\begin{itemize}
	\item Step 1: The PH algorithm is initialized by setting the iteration counter  $i=0$ and PH multiplier $m_s^{(i=0)}=0$. 
	\item Step 2: Each of $N^{\mathrm{s}}$ subproblems is solved in parallel to obtain binary decisions  $u_{kbts}^{(i=0)}$. When all subproblems are solved, we compute $\overline{u}$ as the weighted average of all subproblem solutions. 
	\item Step 3: The iteration counter is updated. For each subproblem we update the value of PH multiplier  $m_s^{(i)}$ using the optimal solution of the previous iteration 	$u_{kbts}^{(i-1)}$, the value of $\overline{u}$ and exogenous penalty coefficient $\rho$.  Then we resolve each subproblem, where the deviation of $u_{kbts}^{(i)}$ from $\overline{u}$ is penalized. After the subproblems are solved, the value of $\overline{u}$ is recomputed using $u_{kbts}^{(i)}$. The iterative process continues until the mismatch $g^{(i)}$ is below a given tolerance $\varepsilon$.
	\end{itemize}
To avoid issues with convergence, we follow the recommendation of  \cite{Watson:2010dn} and set the value of $\rho$	based on the cost coefficients of respective  hedged variables. 
	
		

		
		\begin{algorithm}[tbh]
			\DontPrintSemicolon
			\SetAlgoLined
			Step 1. $i:=0, m_s^{(i=0)}:=0$ \;
 			Step 2. \For{$s\gets1$ \KwTo $N_{s}$}{
 				$u_{kbts}^{(i=0)}\gets\text{arg}\!\min\limits_{\!\!\!\!\!\!\!\!\!\!\!\!u} \gamma\!\cdot\!IC+OC_{s}$\;
 			}
 			$\overline{u}=\sum\limits_{s\in{\cal S}}\omega_s\cdot u_{k,b,t=1,s}^{(i=0)}$\;
 			Step 3. \Do{$\mathrm{convergence\!:}$ $g^{(i)}<\epsilon$}{
				$i\gets i+1$\;
				\For{$s\gets1$ \KwTo $N_{s}$}{
					$m_s^{(i)}\!\gets\!m_s^{(i-1)}+\rho\!\cdot\!\left(u_{k,b,t=1,s}^{(i-1)}\!-\!\overline{u}\right)$\;
					$u_{kbts}^{(i)}\!\gets\!\text{arg}\!\min\limits_{\!\!\!\!\!\!\!\!\!\!\!\!u} \gamma\!\cdot\!IC+OC_{s}+m_s^{(i)}\!\cdot \! u_{k,b,t=1,s}
					+\frac{\rho}{2}\!\left\|{u_{k,b,t=1,s}}\!-\!\overline{u} \right\| ^2 \!\! $\;								
				}
				$\overline{u}=\sum\limits_{s\in{\cal S}}\omega_s\cdot u_{k,b,t=1,s}^{(i)}$\;
				$g^{(i)}\gets \sum\limits_{s\in{\cal S}} \omega_s \!\cdot\! \left\|u_{k,b,t=1,s}^{(i)}-\overline{u}\right\|$\;
 			}
 			\Return $u_{kbts}^{(i)}$
 			\caption{PHA for Installing Mobile ES units}
 			\label{alg:PH_algorithm}
		\end{algorithm}


	\section{Case Study}\label{sec:case}

		The case study is performed based on the 15-bus radial distribution test system described in \cite{Papavasiliou:2017ek}.  The distribution system operator aims to install one mobile ES units with $\overline{E}_k$=1MWh, $\overline{P}_k$=0.15MW and $\aleph^{\mathrm{ch}}$=$\aleph^{\mathrm{dis}}$=0.9.  Based on \cite{pandvzic2015near}, the capital costs are $C^{\mathrm{P}}$=\$1,000/kW and $C^{\mathrm{E}}$=\$50/kWh and the expected lifetime is 10 years. The disaster forecasts are represented by the scenarios described in Table \ref{table:scen_configure}. Case 1 considers the disaster represented by one scenario ($s_2$) and Case 2 considers the disaster  represented by 5 scenarios ($s_2, \cdots ,s_6$). In both cases, the normal operations are represented by one scenario ($s_1$). Each disaster scenario is assumed to start at $t$=6 hour and affect the capacity of the  distribution lines, given in Table~\ref{table:scen_configure}, via parameter $\alpha_{lts}$ during the rest of the time horizon. 		The transition time is defined as $T_{b_1,b_2,t}^{\mathrm{d}}\!=\!\min(\vert b_1-b_2 \vert, d_{b_1,b_2})$ where $d_{b_1,b_2}$ is the number of lines between buses $b_1$ and $b_2$.
		The value of lost load  is $C_b^{\mathrm{VoLL}}$ = \$5,000/MWh. All simulations have been carried out using Gurobi solver v7.5.1 on Julia 0.6 \cite{lubin2015computing} with an Intel Core i7 2.6GHz processor with 16GB of memory.



	
	


		\begin{table}[!b]
			\centering
			\captionsetup{justification=centering, labelsep=period, font=footnotesize, textfont=sc}
			\caption{Disaster Scenarios for Case 1 and Case 2}			
			\begin{tabular}{@{\extracolsep{\fill} } c@{\extracolsep{\fill} } | c  c | c  c  c  c  c  c @{\extracolsep{\fill} }}
				\hline
				\hline
				&\multicolumn{2}{c|}{Case 1}&\multicolumn{6}{c}{Case 2}\\
				\hline
				Scenario $s$& 1&2&1&2&3&4&5&6\\
				\hline
				Probability $\omega_s$ &0.9&0.1&0.9&0.02&0.02&0.02&0.02&0.02\\						
				\hline				
			Line$^*$ $l$  &-&4&-& 1& 12 &4 &8 &9 \\		
				\hline
				\hline
				\multicolumn{9}{l}{$^*$ Line numbers are assigned for the edge connecting bus $b$ with its} \\
				\multicolumn{9}{l}{parent bus as illustrated in Fig.~\ref{fig:transit_2scen}, i.e. $l=b$.}
			\end{tabular}
			\label{table:scen_configure}			
		\end{table}

\subsubsection{Computational performance}\label{subsec:computation_performance}
Table~\ref{table:BFvsPH} compares the proposed PH implementation and the brute-force (BF) implementation (i.e. solving the proposed optimization using Gurobi without any algorithmic enhancement) in terms of their computational performance and optimality. In both cases, the PH and BF implementations attain nearly identical values of the objective functions and respective minimizers. On the other hand, the PH implementation solves the proposed optimization 4x times faster in Case 1, where only 2 scenarios are considered. As the number of scenarios increases to size as in Case 2, the computational gains increase to 8x times. This comparison reveals the computational efficiency of the PH implementation for solving instances with a larger number of scenarios.  Given the superior performance of the PH algorithm described in Table~\ref{table:BFvsPH}, it is used to obtain all numerical results in the rest of the  paper. 

		\begin{table}[!t]
			\centering
			\captionsetup{justification=centering, labelsep=period, font=footnotesize, textfont=sc}
			\caption{Objective Function Values and CPU Times for the BF and PH Implementation}			
			\begin{tabular}{ c | c | c | c | c}
				\hline
				\hline
			\multirow{2}{*}{}&\multicolumn{2}{c|}{Objective function, \$}&\multicolumn{2}{c}{CPU time, s}\\
				\cline{2-5}
				& BF&PH&BF&PH\\				
				\hline
				Case 1&1032.06&1032.07  &35.18&7.99\\
				Case 2 &1787.18&1788.41&21,519&2,594\\
				\hline
				\hline
			\end{tabular}
			\label{table:BFvsPH}			
		\end{table}

\subsubsection{Effect of the ES mobility}\label{subsec:siting_scheduling}

		\begin{table}[!b]
			\centering
			\captionsetup{justification=centering, labelsep=period, font=footnotesize, textfont=sc}
			\caption{Total Lost Load and Objective Function for Instances without ES, with Stationary ES and with Mobile ES}			
			\begin{tabular}{ c | c | c | c | c | c }
				\hline
				\hline
				\multicolumn{1}{c}{}&\multicolumn{2}{c|}{}&w/o ES&Stationary ES&Mobile ES\\
				\hline				
				\multirow{2}{*}{Case}&Total lost load,&$s_1$ &0&0&0\\
				\multirow{2}{*}{1}&\multirow{1}{*}{MWh}&$s_2$ &0.44&0&0\\				
				\cline{2-6}
				&\multicolumn{2}{c|}{Objective function, \$}&1185.62&1032.51&1032.07\\
				
				\hline

				&&$s_1$ &0&0&0\\
				&\multirow{2}{*}{Total}&$s_2$ &3.47&2.89&2.18\\
				\multirow{2}{*}{Case}&\multirow{2}{*}{lost load,}&$s_3$&6.82&6.82&5.42\\
				\multirow{2}{*}{2}&\multirow{2}{*}{MWh}&$s_4$ &0.44&0&0\\
				&&$s_5$ &0&0&0\\
				&&$s_6$ &0&0&0\\
				\cline{2-6}
				&\multicolumn{2}{c|}{Objective function, \$}&2036.20&1997.95&1788.41\\
				\hline
				\hline
			\end{tabular}
			\label{tab:result_lostload_objvalue}			
		\end{table}

		\begin{figure}[!b]
    		\centering
    		\includegraphics[width=\columnwidth]{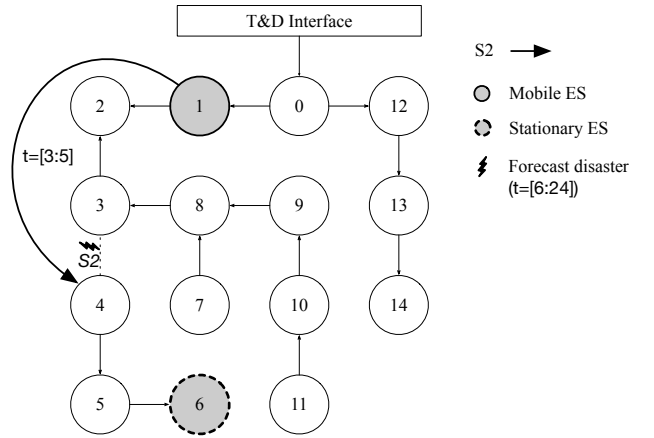}
    		\caption{\small The 15-bus test system from \cite{Papavasiliou:2017ek} used in the case study with the storage placement decisions in the disaster scenario of Case 1.  } 
    		\label{fig:transit_2scen}			
		\end{figure}
		
		\begin{table*}[!t]
			\centering
			\captionsetup{justification=centering, labelsep=period, font=footnotesize, textfont=sc}
			\caption{Bus Locations and State-of-Charge (SoC) of Mobile ES units in Case 2}			
			\begin{tabular}{ @{\extracolsep{\fill} }c@{\extracolsep{\fill} }|@{\extracolsep{\fill} }c@{\extracolsep{\fill} }|@{\extracolsep{\fill} }c@{\extracolsep{\fill} }@{\extracolsep{\fill} }c@{\extracolsep{\fill} }@{\extracolsep{\fill} }c@{\extracolsep{\fill} }@{\extracolsep{\fill} }c@{\extracolsep{\fill} }@{\extracolsep{\fill} }c@{\extracolsep{\fill} }|@{\extracolsep{\fill} }c@{\extracolsep{\fill} }@{\extracolsep{\fill} }c@{\extracolsep{\fill} }@{\extracolsep{\fill} }c@{\extracolsep{\fill} }@{\extracolsep{\fill} }c@{\extracolsep{\fill} }@{\extracolsep{\fill} }c@{\extracolsep{\fill} }|@{\extracolsep{\fill} }c@{\extracolsep{\fill} }@{\extracolsep{\fill} }c@{\extracolsep{\fill} }@{\extracolsep{\fill} }c@{\extracolsep{\fill} }@{\extracolsep{\fill} }c@{\extracolsep{\fill} }@{\extracolsep{\fill} }c@{\extracolsep{\fill} }|@{\extracolsep{\fill} }c@{\extracolsep{\fill} }@{\extracolsep{\fill} }c@{\extracolsep{\fill} }@{\extracolsep{\fill} }c@{\extracolsep{\fill} }@{\extracolsep{\fill} }c@{\extracolsep{\fill} }@{\extracolsep{\fill} }c@{\extracolsep{\fill} }|@{\extracolsep{\fill} }c@{\extracolsep{\fill} }@{\extracolsep{\fill} }c@{\extracolsep{\fill} }@{\extracolsep{\fill} }c@{\extracolsep{\fill} }@{\extracolsep{\fill} }c@{\extracolsep{\fill} }}
			
				\hline
				\hline
				\centering
				&\multirow{2}{*}{$s$}
				&\multicolumn{24}{c}{Time interval \#} \\
				\cline{3-26}
	 			&&1&2&3&4&5&6&7&8&9&10&11&12&13&14&15&16&17&18&19&20&21&22&23&24 \\
	 			\hline	 				 				 			
					&1&1&1&1&1&1&1&1&1&1&1&1&1&1&1&1&1&1&1&1&1&1&1&1&1\\
					&2&1&T&2&2&2&T&1&1&1&1&1&T&0&0&0&T&1&1&1&1&1&T&0&0\\
				  \multirow{2}{*}{Bus \# }  &3&1&T&0&0&T&12&12&12&12&12&12&12&T&0&0&0&0&0&0&T&12&12&12&12\\
					&4&1&T&2&T&T&4&4&4&4&4&4&4&4&4&4&4&4&4&4&4&4&4&4&4\\
					&5&1&1&T&2&T&T&8&8&8&8&8&8&T&T&T&1&1&1&1&1&1&1&--&2\\
					&6&1&T&T&T&T&9&9&9&9&9&T&T&T&T&1&1&1&1&1&1&1&1&1&1\\
				\hline
					&1&0.5&0.5&0.5&0.5&0.5&0.5&0.49&0.49&0.47&0.45&0.42&0.4&0.38&0.38&0.38&0.35&0.33&0.33&0.3&0.3&0.3&0.3&0.3&0.3\\
					&2&0.54&T&0.69&0.84&1&T&0.9&0.75&0.6&0.45&0.3&T&0.45&0.6&0.75&T&0.6&0.45&0.3&0.15&0&T&0.15&0.3\\
					\multirow{2}{*}{SoC, MWh}&3&0.65&T&0.8&0.95&T&0.82&0.68&0.55&0.41&0.27&0.14&0&T&0.15&0.3&0.45&0.6&0.75&0.9&T&0.75&0.6&0.45&0.3\\
					&4&0.64&T&0.79&T&T&0.78&0.76&0.74&0.71&0.68&0.65&0.62&0.59&0.56&0.53&0.5&0.47&0.44&0.41&0.38&0.35&0.33&0.31&0.3\\
					&5&0.35&0.2&T&0.14&T&T&0.27&0.41&0.56&0.7&0.85&1&T&T&T&0.91&0.79&0.67&0.56&0.45&0.37&0.32&T&0.3\\
					&6&0.35&T&T&T&T&0.49&0.62&0.75&0.87&1&T&T&T&T&0.9&0.8&0.8&0.67&0.54&0.43&0.34&0.3&0.3&0.3\\
				\hline
				\hline
				\multicolumn{20}{l}{Labels `T' defines that the mobile ES unit is in transit.}
			\end{tabular}
			\label{tab:result_6scen}
		\end{table*}

		The benefits of mobile ES units can be revealed by comparing their operations with the case with stationary ES units and without ES units. This comparison is given in Table~\ref{tab:result_lostload_objvalue} for Case 1 and Case 2. The effect of the mobile ES units is rather negligible on the objective function in Case 1 due to the small scenario set. However, the use of either stationary or mobile ES units reduces the total lost load (computed as $\sum_{b \in \mathcal{B}} \sum_{t \in \mathcal{T}} p^{\text{ls}}_{bts}$) relative to the case without ES units. On the other hand, the effect of mobile ES units is evident in the numerical results for Case 2. First, using mobile ES units reduces the total lost load across all scenarios considered relative to the simulations  without ES units and with stationary ES units. Second, the mobile ES units ensure the least-cost operation,  achieved in part due to the reduced total load shedding. 	If the ES unit were stationary, as in  \cite{pandvzic2015near} and \cite{7587806}, it would be installed at bus $6$. this decision changes for mobile ES units. Fig.~\ref{fig:transit_2scen} describes the ES placement in Case 1. In this case, the mobile ES is placed at bus $1$ during normal operations and moves to bus $4$ in anticipation of the disaster, which is to occur at $t=6$ hour and to affect line 4. Following the disaster,  the mobile ES unit begins to supply power to downstream buses 5 and 6. The stored energy is enough to avoid load shedding until period $t=24$ hours.  In Case 2, which considers  5 disaster scenarios, the mobile ES is routed and dispatched as described in Table~\ref{tab:result_6scen}. Similarly to Case 1, the mobile ES unit is placed at bus 1 during normal operations ($s_1$) and then has five unique routing trajectories for each disaster scenario ($s_2 \cdots s_6$). These trajectories vary in terms of the optimal locations and dispatch decisions on mobile ES units, which underpins the importance of carefully calibrating disaster scenarios (see \cite{rappaport2009advances}).

	\section{Conclusion}\label{sec:conclusion}

	This paper describes an approach to optimize investments of the distribution system operator in mobile ES units. The ability of mobile ES units to move between different locations is used to trade-off the least-cost operations during normal operations and the need to enhance power grid resilience in case of natural disasters. The proposed optimization is a two-stage mixed-integer program with binary recourse decisions, which account for the relocation of mobile ES units under specific disaster scenarios. The proposed optimization is solved using the PH algorithm. The numerical experiments reveal that the mobile ES reduce the operating costs and the total amount of load shedding caused by natural disasters relative to the cases without ES units or with stationary ES units. 

	\bibliographystyle{IEEEtran}
	\bibliography{PES_GM_2018_v1.bbl}
\end{document}